\documentclass[graybox]{svmult}

\usepackage{mathptmx}       % selects Times Roman as basic font
\usepackage{helvet}         % selects Helvetica as sans-serif font
\usepackage{courier}        % selects Courier as typewriter font
\usepackage{type1cm}        % activate if the above 3 fonts are
\usepackage{makeidx}         % allows index generation
\usepackage{graphicx}        % standard LaTeX graphics tool
\usepackage{multicol}        % used for the two-column index
\usepackage[bottom]{footmisc}% places footnotes at page bottom

\usepackage{subfigure}

\usepackage{amsmath, amssymb}

\newtheorem{rem}{Remark}

\newcommand{\R}{\mathbb{R}}
\newcommand{\Z}{\mathbb{Z}}

\newcommand{\Th}{\mathcal{T}}
\newcommand{\Fh}{\mathcal{F}}

\newcommand{\ip}{{i+\frac12}}
\newcommand{\im}{{i-\frac12}}
\newcommand{\xim}{x_{i-\frac12}}
\newcommand{\xip}{x_{i+\frac12}}

\newcommand{\dx}{\Delta x}
\newcommand{\dt}{\Delta t}

\makeatletter
\newcommand{\sech}{\mathop{\operator@font sech}}
\newcommand{\sign}{\mathop{\operator@font sign}}
\makeatother

\begin{document}

\title*{Dispersive wave runup on non-uniform shores}
\author{Denys Dutykh and Theodoros Katsaounis and Dimitrios Mitsotakis}
\authorrunning{D. Dutykh, Th. Katsaounis and D. Mitsotakis}
\institute{Denys Dutykh \at LAMA, UMR 5127 CNRS, Universit\'e de Savoie, Campus Scientifique, 73376 Le Bourget-du-Lac Cedex, France, \email{Denys.Dutykh@univ-savoie.fr}
\and Theodoros Katsaounis \at Department of Applied Mathematics, University of Crete, Heraklion, 71409 Greece Inst. of App. and Comp. Math. (IACM), FORTH, Heraklion, 71110, Greece \email{thodoros@tem.uoc.gr}
\and Dimitrios Mitsotakis \at IMA, University of Minnesota, Minneapolis MN 55455, USA \email{dmitsot@gmail.com}}

\maketitle

\abstract*{Historically the finite volume methods have been developed for the numerical integration of conservation laws. In this study we present some recent results on the application of such schemes to dispersive PDEs. Namely, we solve numerically a representative of Boussinesq type equations in view of important applications to the coastal hydrodynamics. Numerical results of the runup of a moderate wave onto a non-uniform beach are presented along with great lines of the employed numerical method (see D.~Dutykh \emph{et al.} (2011) \cite{Dutykh2010} for more details).
}

\abstract{Historically the finite volume methods have been developed for the numerical integration of conservation laws. In this study we present some recent results on the application of such schemes to dispersive PDEs. Namely, we solve numerically a representative of Boussinesq type equations in view of important applications to the coastal hydrodynamics. Numerical results of the runup of a moderate wave onto a non-uniform beach are presented along with great lines of the employed numerical method (see D.~Dutykh \emph{et al.} (2011) \cite{Dutykh2010} for more details).
}

\section{Introduction}

The simulation of water waves in realistic and complex environments is a very challenging problem. Most of the applications arise from the areas of coastal and naval engineering, but also from natural hazards assessment. These applications may require the computation of the wave generation \cite{Dutykh2006, Kervella2007}, propagation \cite{Titov1997}, interaction with solid bodies, the computation of long wave runup \cite{TS94, TS} and even the extraction of the wave energy \cite{Simon1981}. Issues like wave breaking, robustness of the numerical algorithm in wet-dry processes along with the validity of the mathematical models in the near-shore zone are some  basic problems in this direction \cite{Hibberd1979}. During past several decades the classical Nonlinear Shallow Water Equations (NSWE) have been essentially employed to face these problems \cite{Dutykh2009a}. Mathematically, these equations represent a system of conservation laws describing the propagation of infinitely long waves with a hydrostatic pressure assumption. The wave breaking phenomenon is commonly assimilated to the formation of shock waves (or hydraulic jumps) which is a common feature of hyperbolic PDEs. Consequently, the finite volume (FV) method has become the method of choice for these problems due to its excellent intrinsic conservative and shock-capturing properties \cite{DeKa, Dutykh2009a}.

In the present article we report on recent results concerning the extension of the finite volume method to dispersive wave equations steming essentially from water wave modeling \cite{Peregrine1967, Dutykh2007, Dutykh2010}.

\section{Mathematical model and numerical methods}

Consider a cartesian coordinate system in two space dimensions $(x,z)$ to simplify notations. The $z$-axis is taken vertically upwards and the $x$-axis is horizontal and coincides traditionally with the still water level. The fluid domain is bounded below by the bottom $z = -h(x)$ and above by the free surface $z = \eta (x,t)$. Below we will also need the total water depth $H(x,t) := h(x) + \eta(x,t)$. The flow is supposed to be incompressible and the fluid is inviscid. An additional assumption of the flow irrotationality is made as well.

In the pioneering work of D.H.~Peregrine (1967) \cite{Peregrine1967} the following system of Boussinesq type equations has been derived:
\begin{equation}\label{eq:p1}
  \eta_t + \bigl((h+\eta)u\bigr)_x = 0,
\end{equation}
\begin{equation}\label{eq:p2}
  u_t + uu_x + g\eta_x - \frac{h}{2} (hu)_{xxt} + \frac{h^2}{6}u_{xxt} = 0,
\end{equation}
where $u(x,t)$ is the depth averaged fluid velocity, $g$ is the gravity acceleration and underscripts ($u_x$, $\eta_t$) denote partial derivatives.

In our recent study \cite{Dutykh2010} we proposed an improved version of this system which contains higher order nonlinear terms which should be neglected from asymptotic point of view and can be written in conservative variables $(H, Q) = (H, Hu)$ as:
\begin{equation}\label{eq:cons1}
  H_t + Q_x = 0,
\end{equation}
\begin{equation}\label{eq:cons2}
\Bigl(\bigl(1+\frac13 H_x^2 - \frac16 HH_{xx}\bigr)Q_t - \frac13 H^2Q_{xxt} - \frac13 HH_xQ_{xt}\Bigr) 
+ \Bigl(\frac{Q^2}{H} + \frac{g}{2}H^2\Bigr)_x = gH h_x.
\end{equation}

Obviously the linear characteristics of both systems \eqref{eq:p1}, \eqref{eq:p2} and \eqref{eq:cons1}, \eqref{eq:cons2} coincide since they differ only by nonlinear terms.

However, this modification has several important implications onto structural properties of the obtained system. First of all, the magnitude of the dispersive terms tends to zero when we approach the shoreline $H \to 0$. This property corresponds to our physical representation of the wave shoaling and runup process. On the other hand, the resulting system becomes invariant under vertical translations (subgroup $G_5$ in Theorem 4.2, T.~Benjamin \& P.~Olver (1982) \cite{Benjamin1982}):
\begin{equation}\label{eq:vert}
  z \leftarrow z + d, \quad \eta \leftarrow \eta - d, \quad
  h \leftarrow h + d, \quad u \leftarrow u,
\end{equation}
where $d$ is some constant. This property is straightforward to check since we use only the total water depth variable $H = h + \eta$ which remains invariant under transformation \eqref{eq:vert}.

\begin{rem}
In this paper we will consider the initial-boundary value problem  posed in a bounded domain $I=[b_1,b_2]$ with reflective boundary conditions. In this case one needs to impose boundary conditions only in one of the two dependent variables, cf. \cite{FP}.  In the case of reflective boundary conditions it is sufficient to take $u(b_1,t)=u(b_2,t)=0$. 
\end{rem}

\subsection{Finite volume discretization}

Let $\Th= \{x_i\}, \ i\in\Z$ denotes a partition of $\R$ into cells $C_i= (\xim,\xip)$ where $x_i = (x_{\ip}+x_{\im})/2$ denotes the midpoint of $C_i$. Let $\dx_i= \xip-\xim$ be the length of the cell $C_i$,  $\dx_{\ip}=x_{i+1}-x_i$. (Here, we consider only uniform grids with $\Delta x_i=\Delta x_{i+\frac{1}{2}}=\Delta x$.)

The governing equations \eqref{eq:cons1}, \eqref{eq:cons2} can be recast in the following vector form:
\begin{equation*}
[{\bf D}({\bf v_t})] + [{\bf F}({\bf v})]_x = {\bf S}({\bf v}),
\end{equation*}
where
\begin{align} 
& {\bf D}({\bf v_t})=
\begin{pmatrix} H_t \\  (1+\frac{1}{3}H_x^2-\frac{1}{6}HH_{xx})Q_t
-\frac13 H^2Q_{xxt} - \frac{1}{3}HH_x Q_{xt}
\end{pmatrix},  \label{E5.2} \\
&  {\bf F}({\bf v}) =
\begin{pmatrix}
Q\\
\frac{Q^2}{H} + \frac{g}{2}H^2
\end{pmatrix},\qquad 
{\bf S}({\bf v})=\begin{pmatrix} 0\\ gHh_x \end{pmatrix}. \label{E5.3}
\end{align}

We denote by $H_i$ and $U_i$ the corresponding cell averages. To discretize the dispersive terms in \eqref{E5.2} we consider the following approximations:
\begin{multline*}
\frac{1}{\dx}\int_{x_{i-\frac{1}{2}}}^{x_{i+\frac{1}{2}}}\left[1+\frac{1}{3}(H_x)^2-\frac{1}{6}HH_{xx}\right]\;Q\;dx \approx \\
\left(1+\frac{1}{3}\left(\frac{H_{i+1}-H_{i-1}}{2\dx} \right)^2 - \frac{1}{6}H_i \;\frac{H_{i+1}-2H_i+H_{i-1}}{\dx^2} \right) Q_i, 
\end{multline*}

\begin{equation*}
\frac{1}{\dx}\int_{x_{i-\frac{1}{2}}}^{x_{i+\frac{1}{2}}} \frac{1}{3}HH_xQ_x \;dx \approx 
\frac{1}{3}H_i\;\frac{H_{i+1}-H_{i-1}}{2\dx}\frac{Q_{i+1}-Q_{i-1}}{2\dx}, 
\end{equation*}

\begin{equation*}
\frac{1}{\dx}\int_{x_{i-\frac{1}{2}}}^{x_{i+\frac{1}{2}}} \frac{1}{3}H^2 Q_{xx}\;dx \approx 
\frac{1}{3}H_i^2\;\frac{Q_{i+1}-2Q_i+Q_{i-1}}{\dx^2}.
\end{equation*}

We note that we approximate the reflective boundary conditions by taking the cell averages  of  $u$ on the first and the last cell  to be $u_0=u_{N+1}=0$.  We do not impose explicitly boundary conditions on $H$. The reconstructed values on the first and the last cell are computed using neighboring ghost cells and taking odd and even extrapolation for $u$ and $H$ respectively. These specific boundary conditions appeared to reflect incident waves on the boundaries while conserving the mass.
 
This discretization leads to a linear system with tridiagonal matrix denoted by ${\bf L}$ that can be inverted efficiently by a variation of Gauss elimination for tridiagonal systems with computational complexity $O(n)$, $n$-being the dimension of the system.  We note that on the dry cells the matrix becomes diagonal since $H_i$ is zero on dry cells. For the time integration the explicit third-order TVD-RK method is used. In the numerical experiments we observed that the fully discrete scheme is stable and preserves the positivity of $H$ during the runup under a mild restriction on the time step $\dt$.

Therefore, the semidiscrete problem of (\ref{E5.2}) - (\ref{E5.3}) is written as a system of ODEs in the form:
\begin{equation*}
{\bf L}_i{{\bf v}_i}_t+\frac{1}{\dx}({\Fh}_{i+\frac{1}{2}}-{\Fh}_{i-\frac{1}{2}})=\frac{1}{\dx}{\bf S_i}, 
\end{equation*}
where ${\bf L}_i$ is the $i-$th row of matrix ${\bf L}$ and ${\Fh}_{i+\frac{1}{2}}$ can be chosen as one of the numerical flux functions \cite{Dutykh2010} (in computations presented below we choose the FVCF flux \cite{Ghidaglia2001}). In the sequel we will use the KT and the CF numerical fluxes. In this case the Jacobian of ${\bf F}$ is given by the matrix
$$A=\begin{pmatrix} 0 & 1 \\ gH-(Q/H)^2 & 2Q/H \end{pmatrix},$$ and the eigenvalues
are $\lambda_{1,2}=Q/H\pm\sqrt{gH}$.  Therefore, the characteristic numerical flux \cite{Ghidaglia2001} takes the form
\begin{equation*}
{\Fh}_{i+\frac{1}{2}}=\frac{{\bf F}({\bf V}_{i+\frac{1}{2}}^L)+{\bf F}({\bf V}_{i+\frac{1}{2}}^R)}{2}-{\bf U}({\boldsymbol \mu})\frac{{\bf F}({\bf V}_{i+\frac{1}{2}}^R)-{\bf F}({\bf V}_{i+\frac{1}{2}}^L)}{2}, 
\end{equation*}
where ${\boldsymbol \mu}=(\mu_1,\mu_2)^T$ are the Roe average values,
$$
\mu_1=\frac{H_{i+\frac{1}{2}}^L+H_{i+\frac{1}{2}}^R}{2}, \quad 
\mu_2=\frac{\sqrt{H_{i+\frac{1}{2}}^L}U_{i+\frac{1}{2}}^L+\sqrt{H_{i+\frac{1}{2}}^R}U_{i+\frac{1}{2}}^R} {\sqrt{H_{i+\frac{1}{2}}^L}+\sqrt{H_{i+\frac{1}{2}}^R}}
$$ 
 and
\begin{equation*}
{\bf U}({\boldsymbol \mu})=\begin{pmatrix}\frac{s_2(\mu_2+c)-s_1(\mu_2-c)}{2c} & \frac{s_1-s_2}{2c}\\
\frac{(s_2-s_1)(\mu_2^2-c^2)}{2c} & \frac{s_1(\mu_2+c)-s_2(\mu_2-c)}{2c} \end{pmatrix}, \ c=\sqrt{g \mu_1},\ s_i=\sign(\lambda_i) . 
\end{equation*}

For more details on the discretization and reconstruction procedures, (that are based on the hydrostatic reconstraction, \cite{Audusse2004}), we refer to our complete work on this subject \cite{Dutykh2010}.

\section{Numerical results}

In the present section we show a numerical simulation of a solitary wave runup onto a non-uniform sloping beach. More precisely, we add a small pond along the slope. As our results indicate, this small complication is already sufficient to develop some instabilities which remain controlled in our simulations.

As an initial condition we used an approximate solitary wave solution of the following form:
\begin{equation*}
  \eta_0(x) = A_s{\sech}^2\bigl(\lambda (x - x_0)\bigr), \quad
  u_0(x) = -c_s\frac{\eta_0(x)}{1+\eta_0(x)},
\end{equation*}
where $A_s$ is the amplitude relative to the constant water depth taken to be unity in our study. The solitary wave speed $c_s$ along with the wavelength $\lambda$ are given here:
\begin{equation*}
  \lambda=\sqrt{\frac{3A_s}{4(1+A_s)}},, \quad 
 c_s=\sqrt{g}\frac{\sqrt{6}(1+A_s)}{\sqrt{3+2A_s}}\cdot \frac{\sqrt{(1+A_s)\log (1+A_s)-A_s}}{A_s}.
\end{equation*}
The solitary wave is centered initially at $x_0 = 10.62$ and has amplitude $A_s = 0.08$. The constant slope $\beta$ is equal to $2.88^\circ$. The sketch of the computational domain can be found in \cite{Dutykh2010}.

In numerical simulations presented below we used a uniform space discretization with $\dx = 0.025$ and very fine time step $\dt = \dx/100$ to guarantee the accuracy and stability during the whole simulation.

\begin{figure}
  \centering
  \subfigure[$t = 1$ s]{\includegraphics[width=0.45\textwidth]{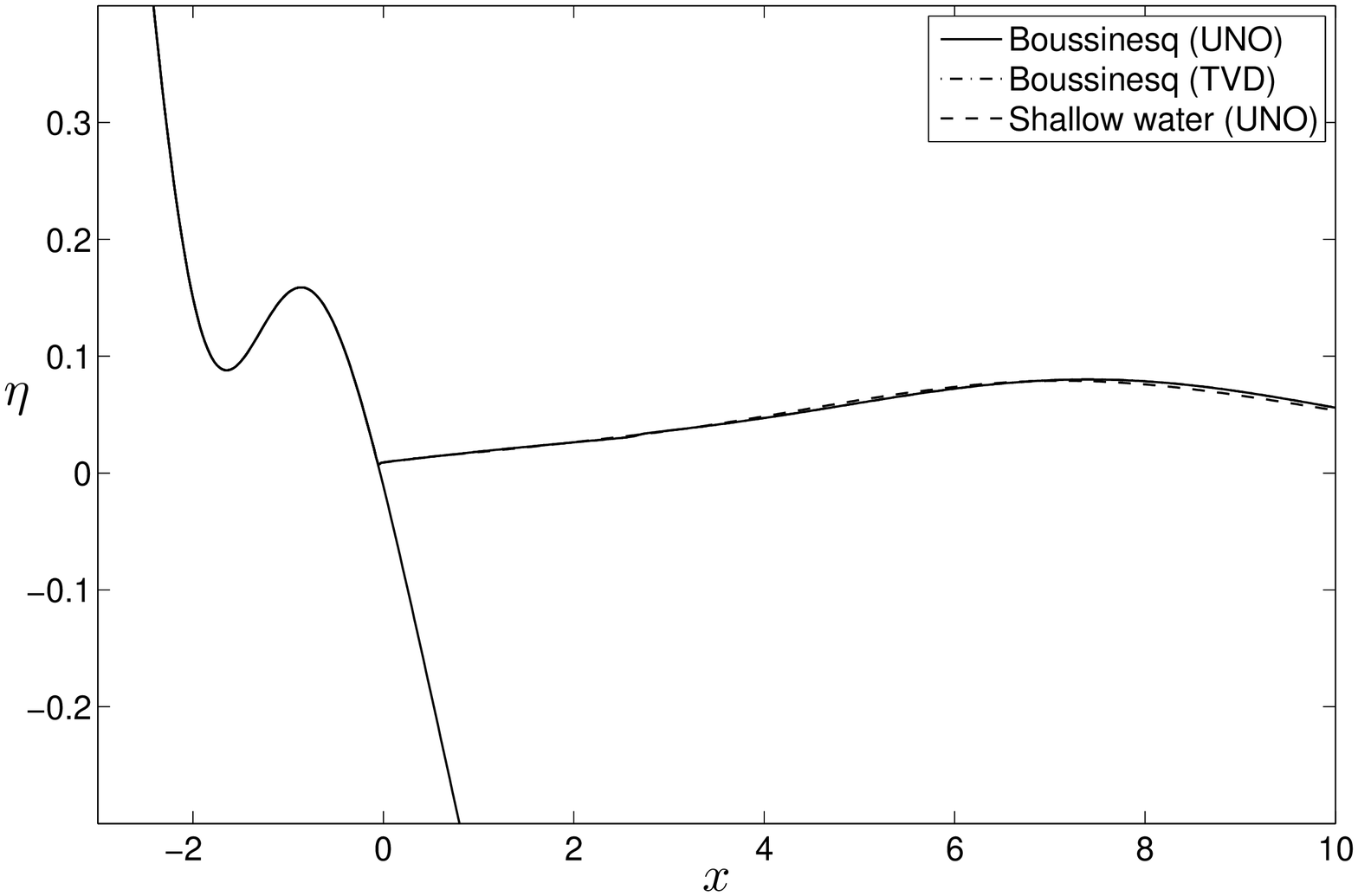}}
  \subfigure[$t = 3$ s]{\includegraphics[width=0.45\textwidth]{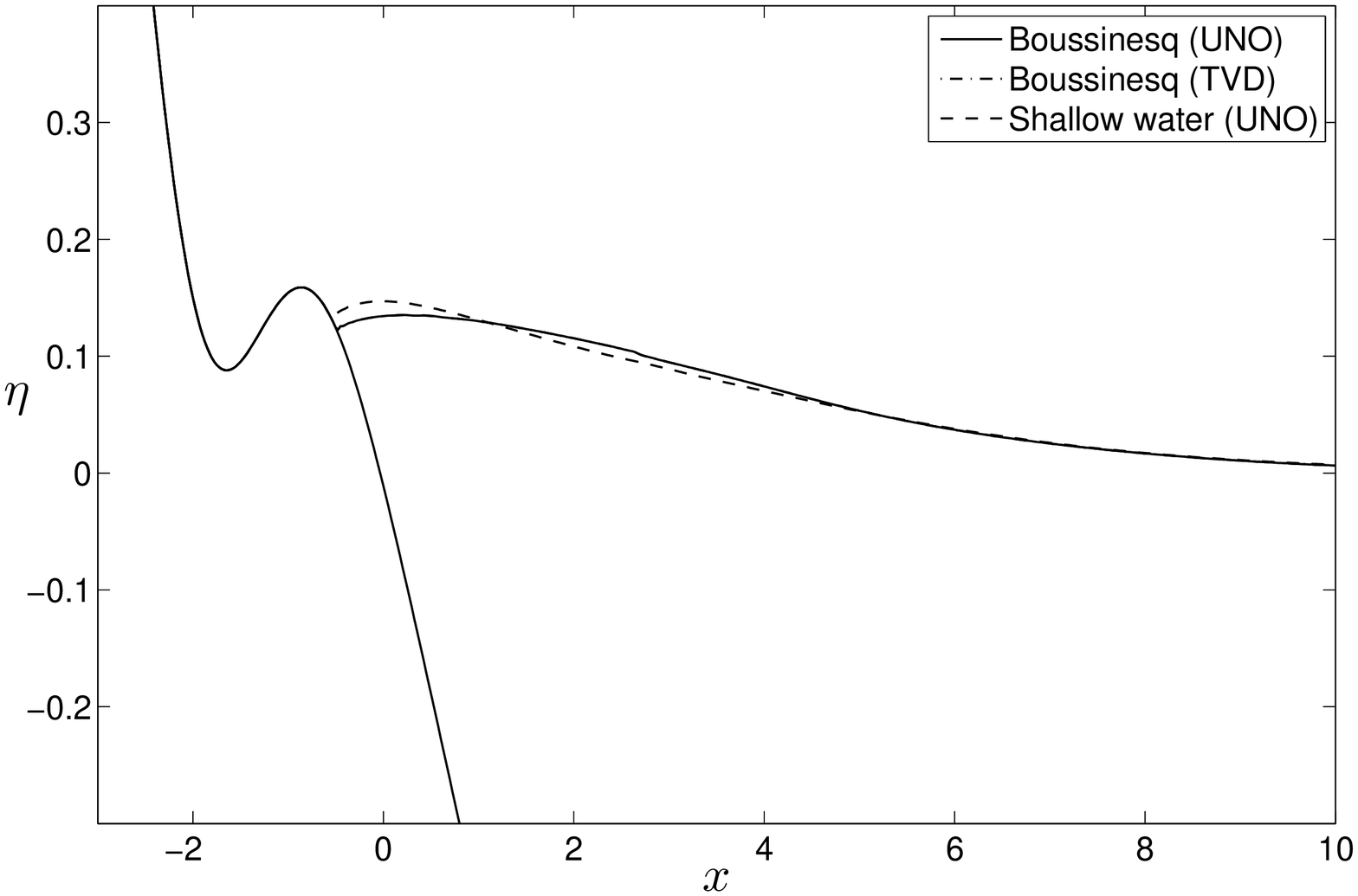}}
  \caption{Solitary wave aproaching a sloping beach with a pond.}
  \label{fig:time13}
\end{figure}

\begin{figure}
  \centering
  \subfigure[$t = 3.5$ s]{\includegraphics[width=0.45\textwidth]{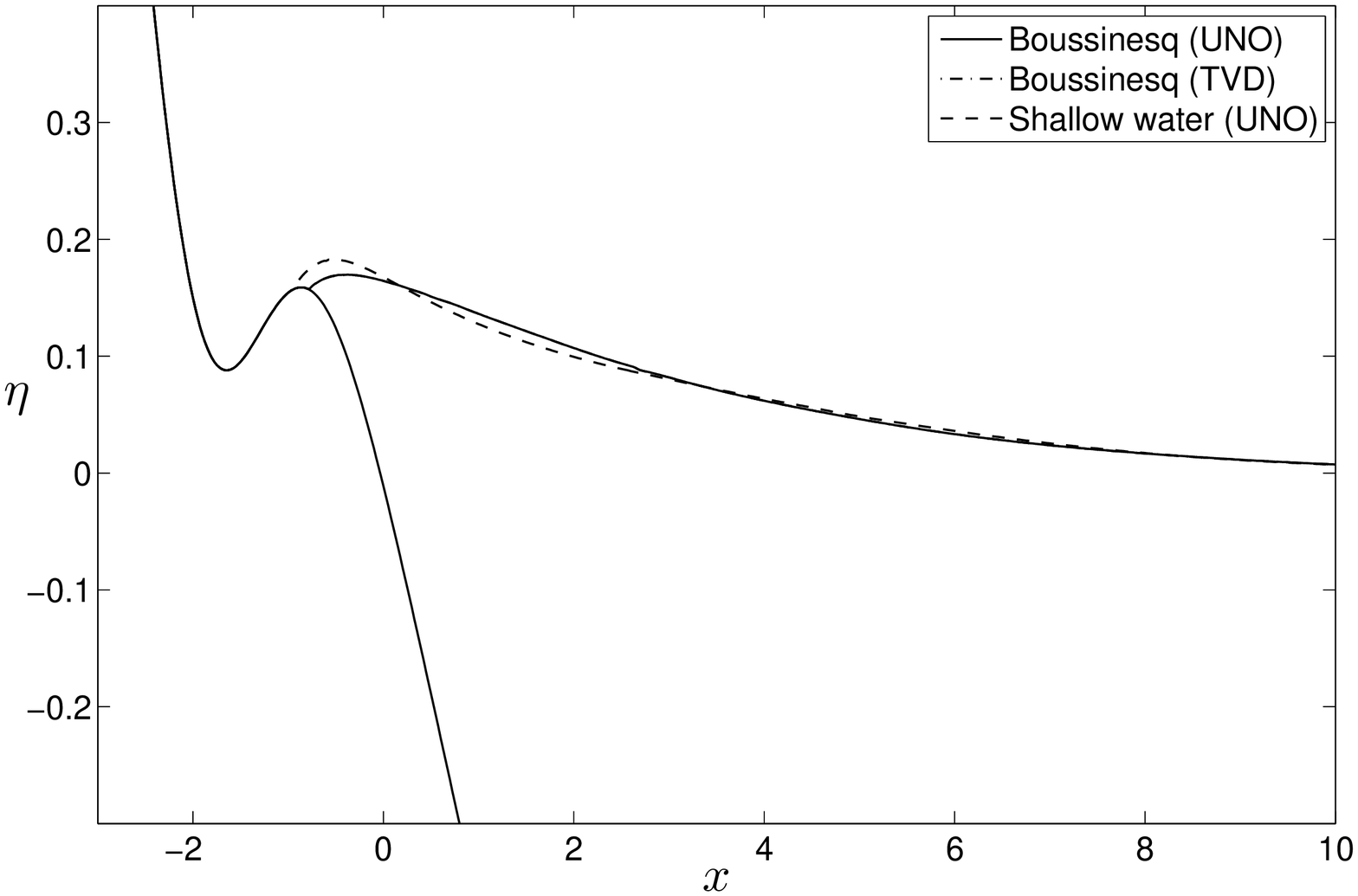}}
  \subfigure[$t = 4$ s]{\includegraphics[width=0.45\textwidth]{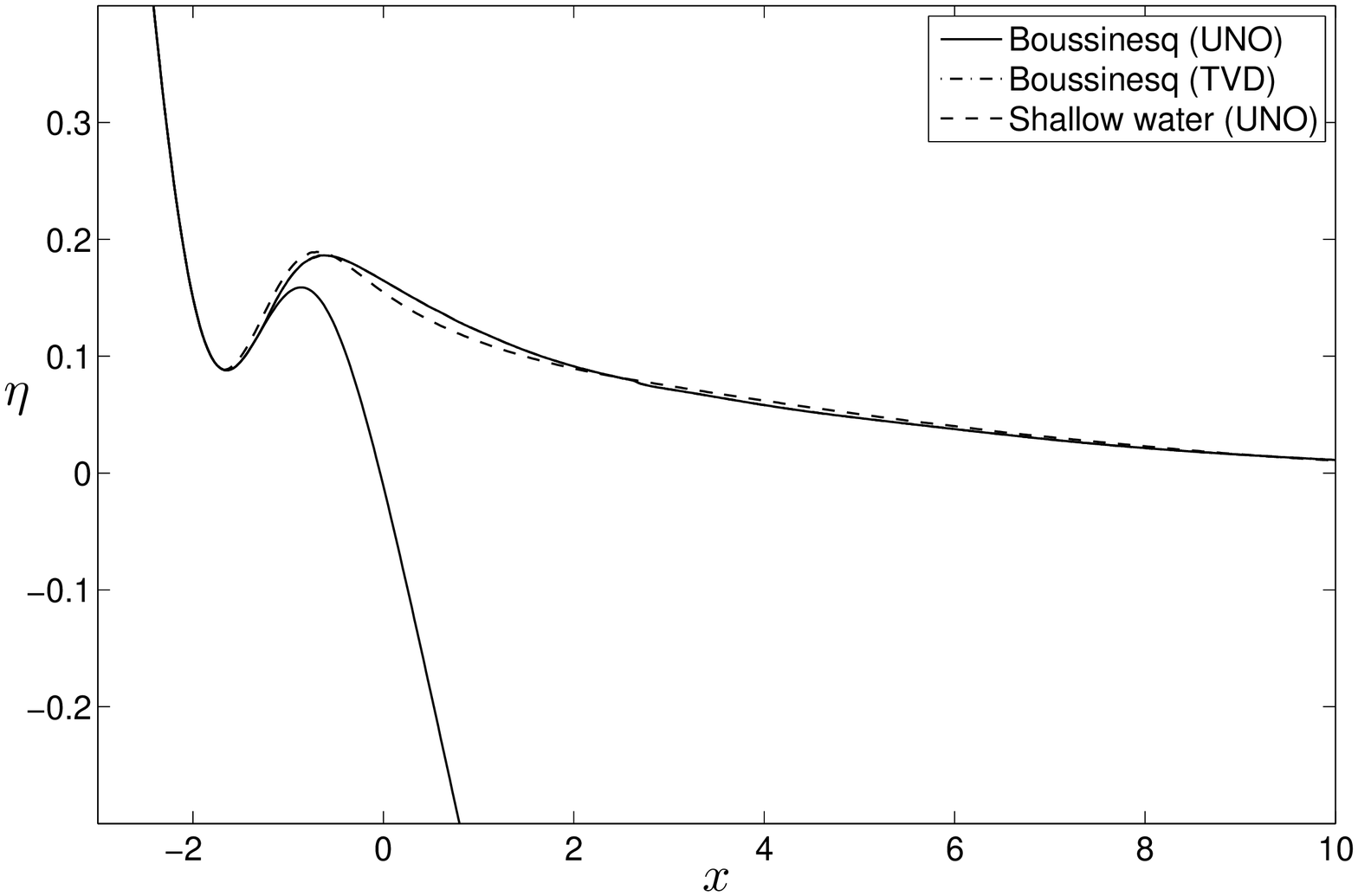}}
  \caption{Beginning of the pond inundation.}
  \label{fig:time34}
\end{figure}

\begin{figure}
  \centering
  \subfigure[$t = 5$ s]{\includegraphics[width=0.45\textwidth]{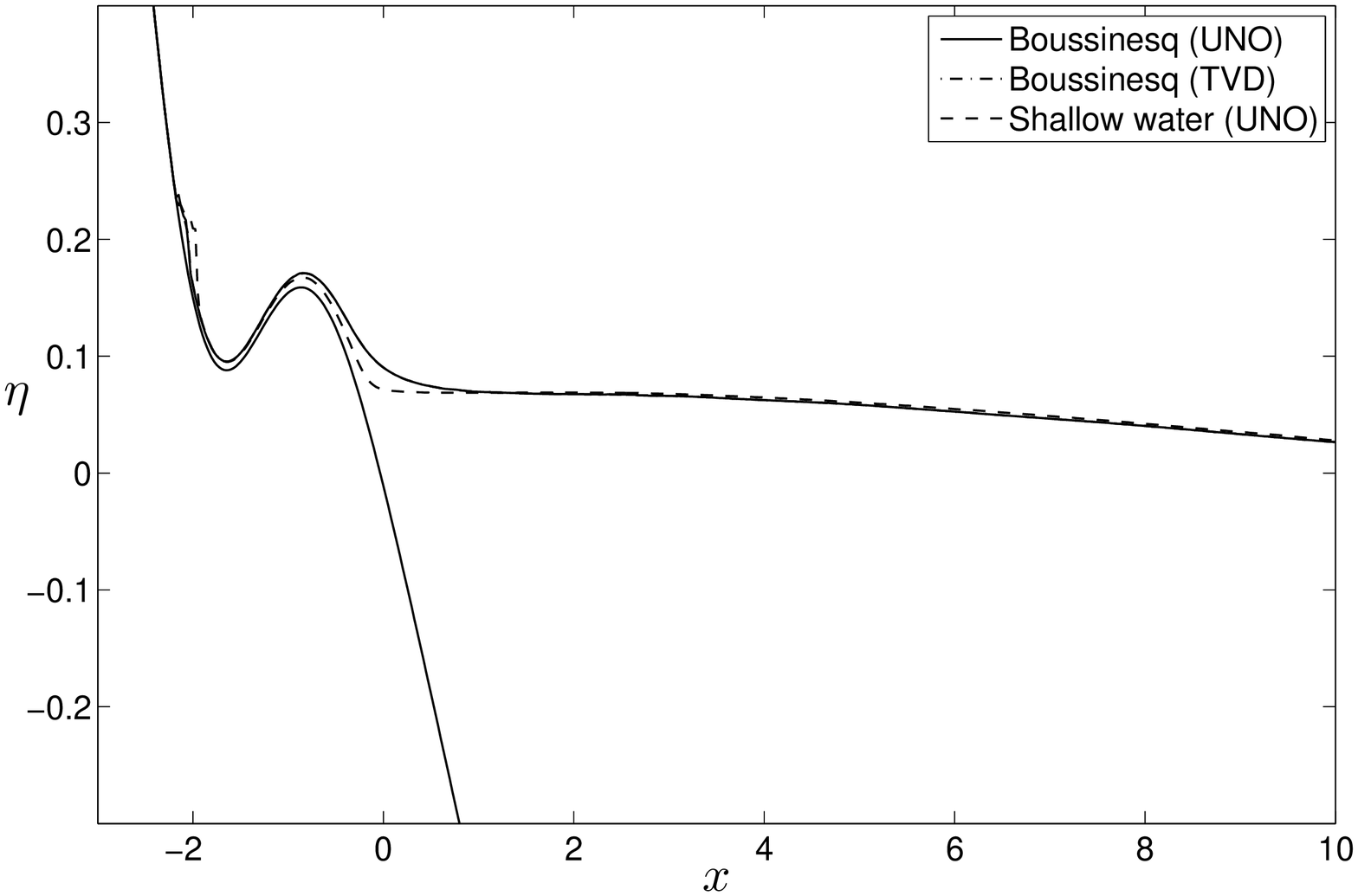}}
  \subfigure[$t = 5.5$ s]{\includegraphics[width=0.45\textwidth]{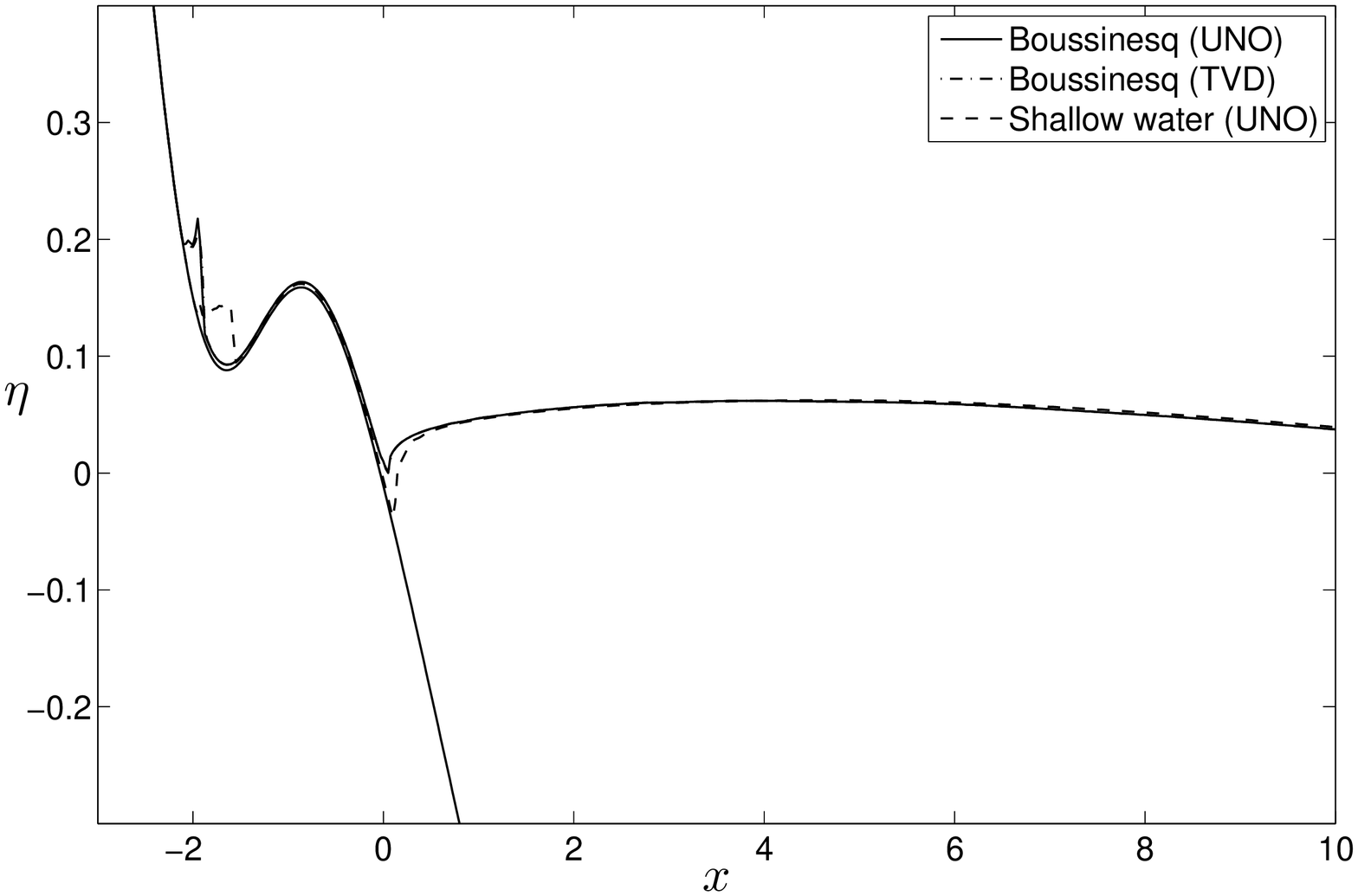}}
  \caption{A part of the wave mass is trapped in the pond volume.}
  \label{fig:time5}
\end{figure}

\begin{figure}
  \centering
  \subfigure[$t = 6$ s]{\includegraphics[width=0.45\textwidth]{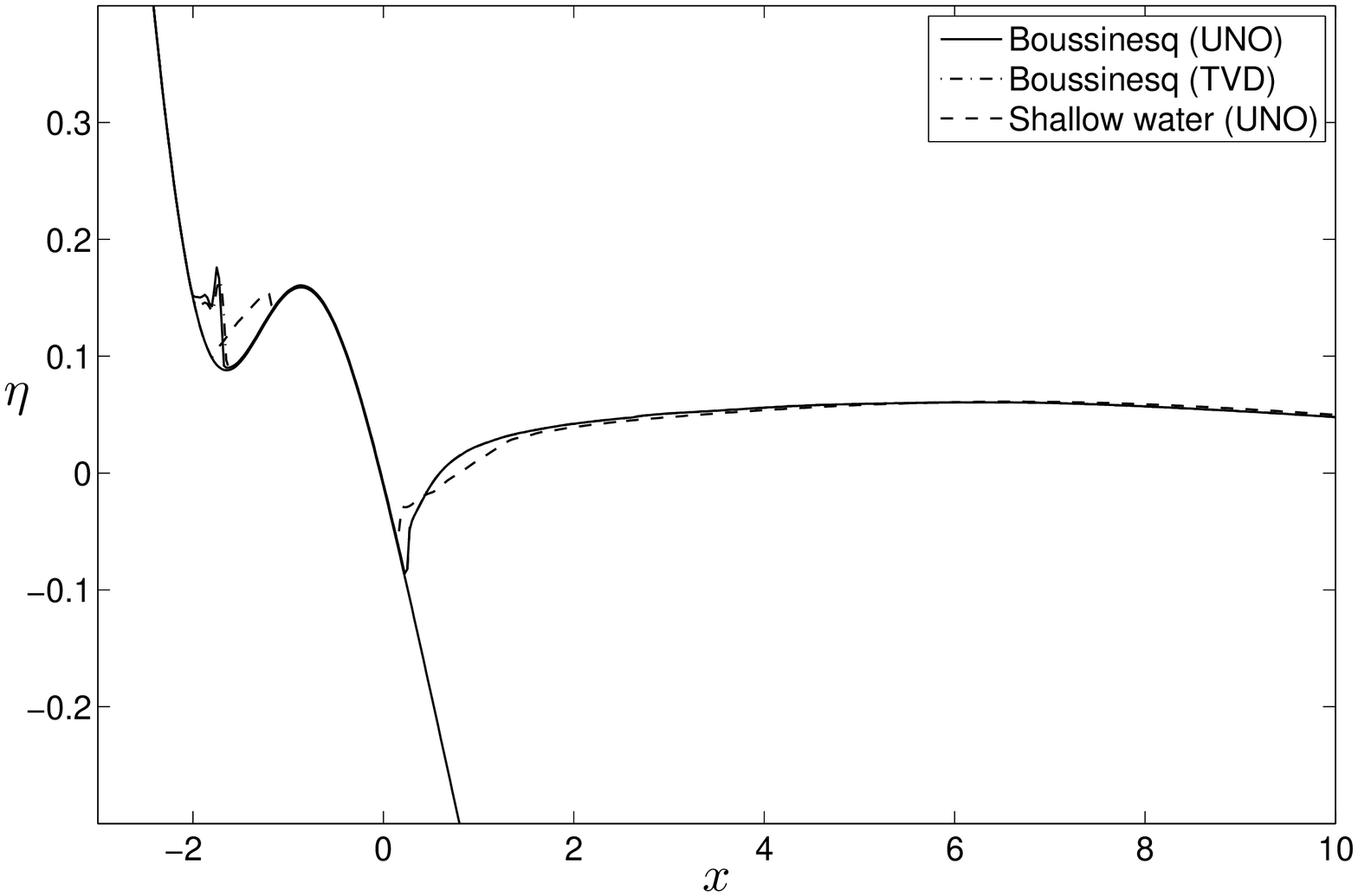}}
  \subfigure[$t = 6.5$ s]{\includegraphics[width=0.45\textwidth]{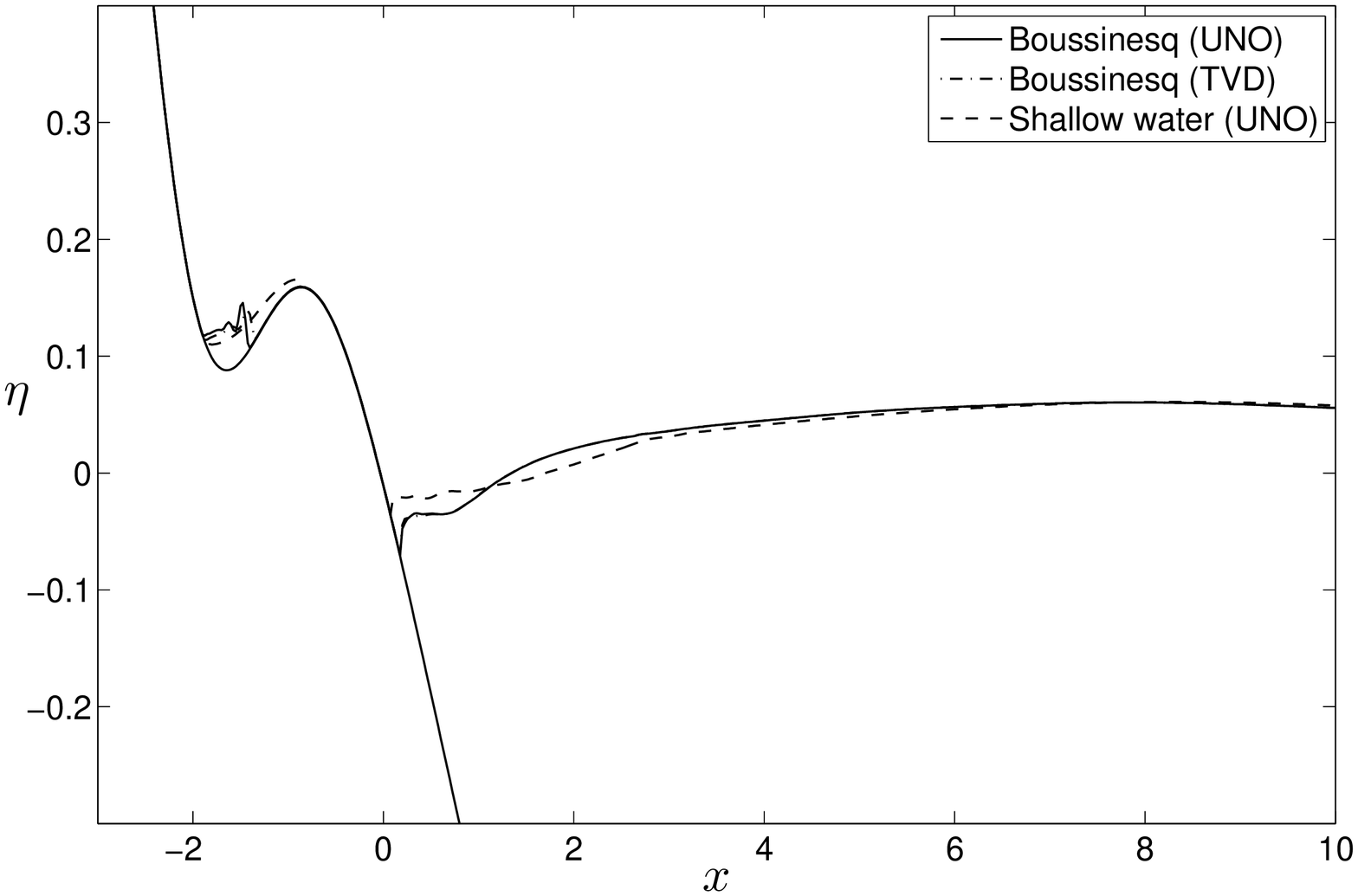}}
  \caption{Wave oscillations in the pond.}
  \label{fig:time6}
\end{figure}

\begin{figure}
  \centering
  \subfigure[$t = 7$ s]{\includegraphics[width=0.45\textwidth]{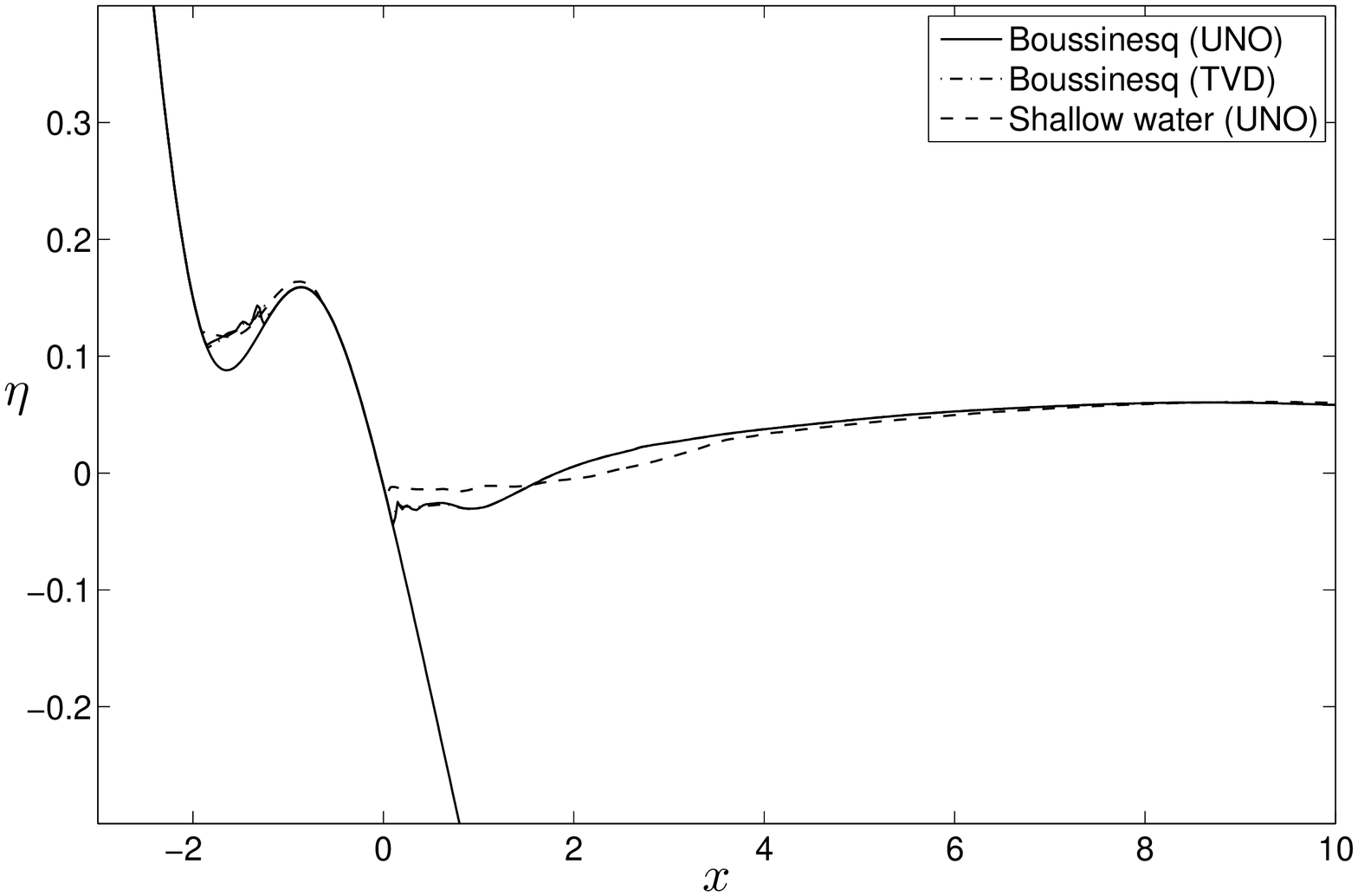}}
  \subfigure[$t = 8$ s]{\includegraphics[width=0.45\textwidth]{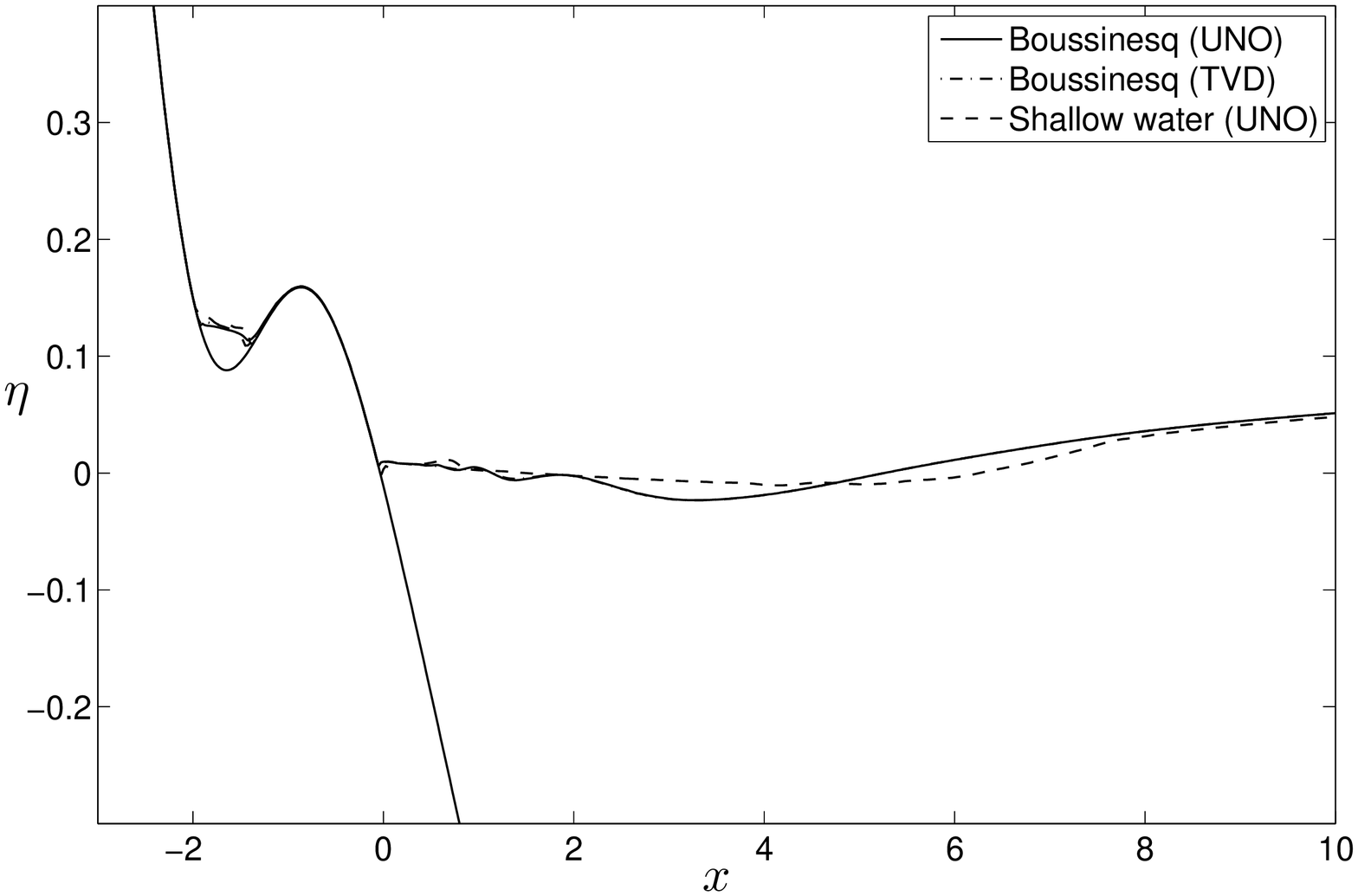}}
  \caption{Stabilization of wave oscillations.}
  \label{fig:time78}
\end{figure}

\begin{figure}
  \centering
  \includegraphics[width=0.5\textwidth]{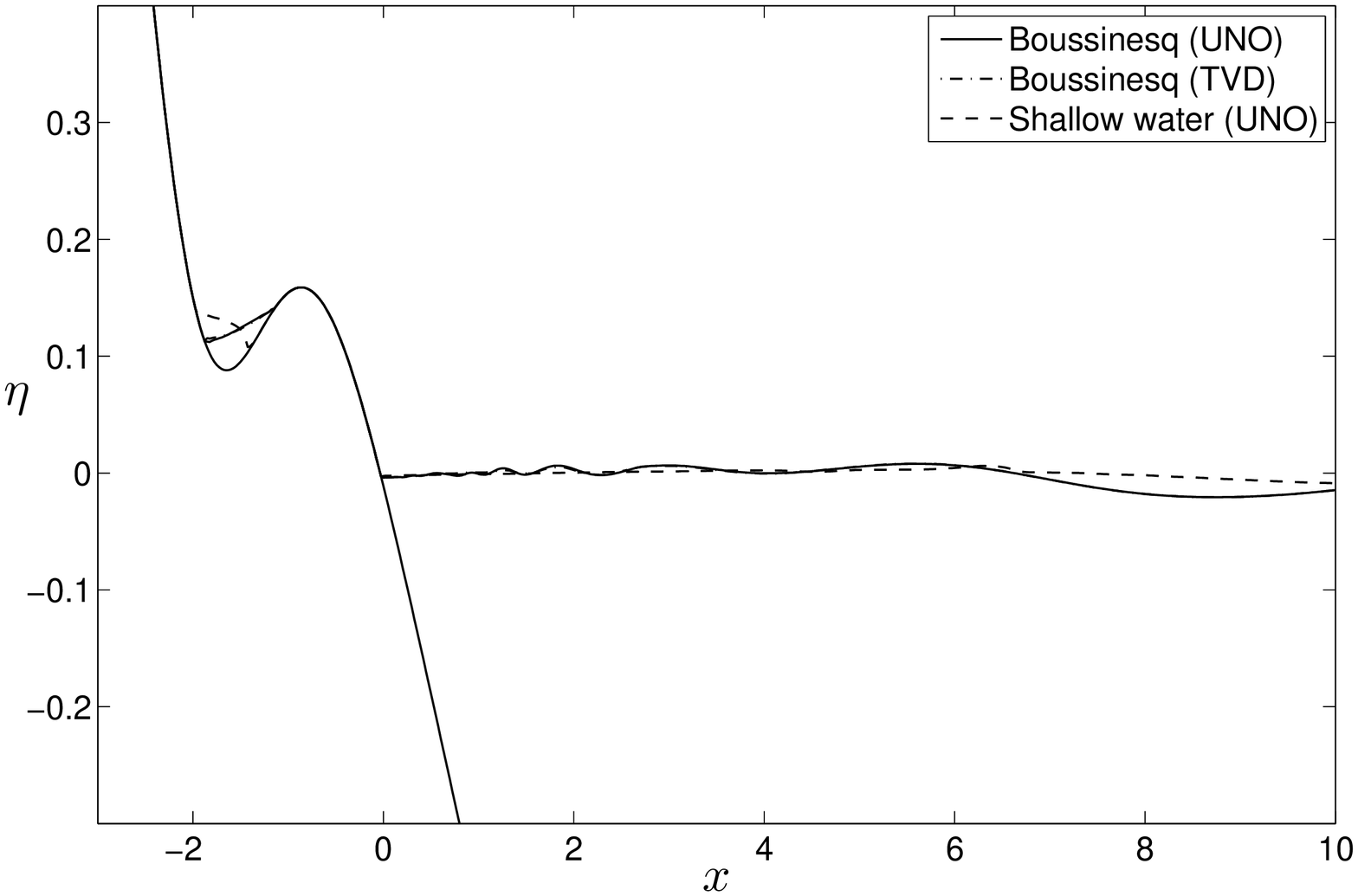}
  \caption{The whole system is tending to the rest position ($t = 10$ s).}
  \label{fig:time10}
\end{figure}

Snapshots of numerical results are presented on Figures \ref{fig:time13} -- \ref{fig:time10}. We present simultaneously three different computational results:
\begin{itemize}
  \item Modified Peregrine system solved with UNO2 reconstruction \cite{HaOs}
  \item The same system with classical MUSCL TVD2 scheme \cite{Leer1979}
  \item Nonlinear Shallow Water Equations (NSWE) with UNO2 scheme \cite{HaOs}
\end{itemize}

Surprisingly good agreement was obtained among all three numerical models. Presumably, the complex runup process under consideration is governed essentially by nonlinearity. However, on Figures \ref{fig:time13}(b) and \ref{fig:time34}(a) the amplitude predicted by NSWE is slightly overestimated.

On Figures \ref{fig:time5}(b) -- \ref{fig:time6}(b) some oscillations (due to the small-dispersion effect characterizing dispersive wave breaking procedures) can be observed. However, their amplitude remains small for all times and does not produce any blow up phenomena. Later these oscillations decay tending gradually to the ``lake at the rest" state (see Figures \ref{fig:time78}, \ref{fig:time10}). 

In the specific experiment a friction term could be beneficial to reduce the amplitude of oscillations (or damp them out completely). However, we prefer to present the computational results of our model without adding any ad-hoc term to show its original performance.

\section{Conclusions}

In this study we presented an improved version of the Peregrine system which is particularly suited for the simulation of dispersive waves runup. This system allows for the description of higher amplitude waves due to improved nonlinear characteristics. Better numerical stability properties have been obtained since most of the dispersive terms tend to zero when we approach the shoreline. Consequently, our model naturally degenerates to classical Nonlinear Shallow Water Equations (NSWE) for which the runup simulation technology is completely mastered nowadays. However we underline that there is no artificial parameter to turn off dispersive terms. Their importance is naturally governed by the underlying physical process.

%Moreover we presented some numerical results on the wave runup onto a complex beach containing a pond. Even in this stiff case the numerical model at hand produced stable and physical results, thus validating modification of the Peregrine system.

\begin{acknowledgement}
D.~Dutykh acknowledges the support from French Agence Nationale de la Recherche, project MathOcean (Grant ANR-08-BLAN-0301-01) and Ulysses Program of the French Ministry of Foreign Affairs under the project 23725ZA. The work of Th.~Katsaounis was partially supported by European Union FP7 program Capacities(Regpot 2009-1), through ACMAC (http://acmac.tem.uoc.gr).
\end{acknowledgement}

%%%% Bibliography  %%%%%%%%%%
\bibliographystyle{spmpsci}
\bibliography{biblio}

\noindent{\small The paper is in final form and no similar paper has been or
  is being submitted elsewhere.}

\end{document}